\documentclass[a4paper]{article}

\usepackage{INTERSPEECH2022}
\usepackage{acronym}
\usepackage{tabularx}
\usepackage{makecell}
\usepackage{multirow}
\usepackage{amssymb}
\usepackage{pifont}
\usepackage{float}

\title{Separator-Transducer-Segmenter: \\ Streaming Recognition and Segmentation of Multi-party Speech}
\name{Ilya Sklyar, Anna Piunova, Christian Osendorfer}
\address{Amazon Alexa}
\email{\{ilsklyar, piunova, osendorf\}@amazon.com}

\begin{document}

\maketitle
%% See more instructions here: 
% http://ctan.math.illinois.edu/macros/latex/contrib/acronym/acronym.pdf

% add new acronyms here
\newacro{AIR}[AIR]{acoustic impulse response}
\newacro{ASR}[ASR]{Automatic speech recognition}
\newacro{cpWER}[cpWER]{concatenated minimum-permutation word error rate}
\newacro{CTC}[CTC]{connectionist temporal classification}
\newacro{DAT}[DAT]{deterministic assignment training}
\newacro{DAT-MS-RNN-T}[DAT-MS-RNN-T]{MS-RNN-T with deterministic assignment training}
\newacro{E2E}[E2E]{end-to-end}
\newacro{ISM}[ISM]{image source method}
\newacro{LSTM}[LSTM]{long short term memory}
\newacro{MS-RNN-T}[MS-RNN-T]{multi-speaker recurrent neural network transducer}
\newacro{MT-MS-RNN-T}[MT-MS-RNN-T]{multi-turn multi-speaker RNN-T}
\newacro{MT-RNN-T}[MT-RNN-T]{multi-turn recurrent neural network transducer}
\newacro{OB}[OB]{overlap-based}
\newacro{OED WER}[OED WER]{optimal edit distance WER}
\newacro{ORC}[ORC]{optimal reference combination}
\newacro{ORC WER}[ORC WER]{optimal reference combination WER}
\newacro{PIT}[PIT]{permutation invariant training}
\newacro{PIT-MS-RNN-T}[PIT-MS-RNN-T]{MS-RNN-T with permutation invariant training}
\newacro{SD}[SD]{speaker-discriminative}
\newacro{RIRs}[RIRs]{room impulse responses}
\newacro{RNN-T}[RNN-T]{recurrent neural network transducer}
\newacro{SB}[SB]{speaker-based}
\newacro{SNR}[SNR]{speech-to-noise ratio}
\newacro{SOT}[SOT]{serialized output training}
\newacro{WER}[WER]{word error rate}
\newacro{WERR}[WERR]{word error rate reduction}
\newacro{SURT}[SURT]{streaming unmixing and recognition transducer}
\newacro{VAD}[VAD]{voice activity detection}
\newacro{STS}[STS]{separator-transducer-segmenter}
\newacro{t-SOT}[t-SOT]{token-level serialized output training}
\newacro{NLU}[NLU]{natural language understanding}
% define the indefinte article for an acronym
\newacroindefinite{ASR}{an}{an}
\newacroindefinite{E2E}{an}{an}
\newacroindefinite{RNN-T}{an}{an}
\begin{abstract}
Streaming recognition and segmentation of multi-party conversations with overlapping speech is crucial for the next generation of voice assistant applications. In this work we address its challenges discovered in the previous work on \ac{MT-RNN-T} with a novel approach, \ac{STS}, that enables tighter integration of speech separation, recognition and segmentation in a single model. First, we propose a new segmentation modeling strategy through \emph{start-of-turn} and \emph{end-of-turn} tokens that improves segmentation without recognition accuracy degradation. Second, we further improve both speech recognition and segmentation accuracy through an emission regularization method, FastEmit, and multi-task training with speech activity information as an additional training signal. Third, we experiment with end-of-turn emission latency penalty to improve end-point detection for each speaker turn. Finally, we establish a novel framework for segmentation analysis of multi-party conversations through emission latency metrics. With our best model, we report 4.6\% abs. turn counting accuracy improvement and 17\% rel. \ac{WER} improvement on LibriCSS dataset compared to the previously published work. 
\end{abstract}
\vspace*{-2mm}
\section{Introduction}
\label{sec:introduction}
\vspace*{-1mm}

% describe motivation behind this work
\ac{ASR} of multi-party recordings with overlapping speech has posed a major scientific challenge for many decades \cite{Hain2007_AMI,fox13b_interspeech,Liu2016_SWC2,Barker_2018}. While single-speaker \ac{ASR} became ubiquitous through applications like voice assistants (Amazon Alexa, Google Home, etc.), its current capability is limited to scenarios with one active speaker at a time. 
%On the other hand, conversation analysis demonstrated that natural human turn-taking dynamics is much more nuanced than one-way voice commands usually addressed to voice assistants, and often involves overlapping speech \cite{sacks}.
Apart from \ac{ASR}, speech overlaps also introduce additional challenges to other parts of the traditional speech processing pipeline as speech segmentation and speaker diarization \cite{PARK2022}.

%% progress made in recent research, from modular systems to joint system gradually
Multi-speaker \ac{ASR} problem was attacked before with both independently optimized modules such as speech separation and speech recognition \cite{Isik_2016, Menne_2019, Settle_2018, von_Neumann_2020, Neumann_2020}, and jointly optimized multi-speaker end-to-end \ac{ASR} systems \cite{Yu_2017_icassp, Yu_2017_interspeech, Qian_2018, Seki_2018, Chang_2019, Tripathi_2020}. In \cite{k2020serialized} a joint multi-speaker \ac{ASR} and speaker change detection system was proposed to tackle speech recognition and segmentation problems simultaneously in the presence of overlapping speech from arbitrary number of speakers. Follow-up work on \ac{SOT} \cite{kanda2020joint, kanda2021largescale, kanda2021slt, kanda2021endtoend, kanda2021comparative} extended it to speaker-attributed \ac{ASR} that can transcribe ``who spoke what" in real multi-speaker conversations with a single integrated model.

\begin{figure}[ht]
	\includegraphics[width=\linewidth]{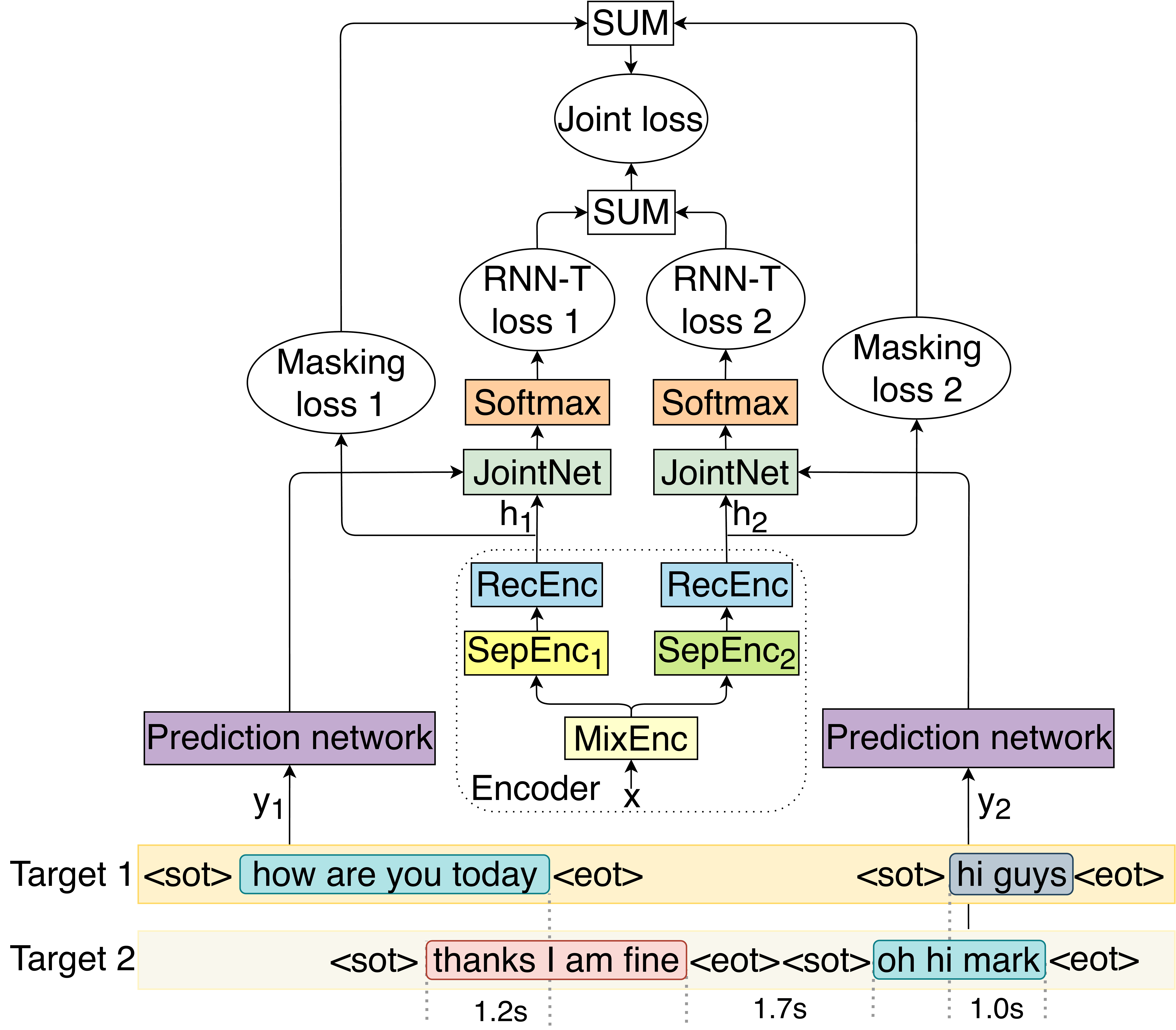}
	\vspace*{0.1mm}
	\vspace*{-6mm}
	\caption{Separator-Transducer-Segmenter. \texttt{<sot>} and \texttt{<eot>} represent start-of-turn and end-of-turn tokens. Model blocks with the same colour have tied parameters, transcripts in the colour-matched boxes belong to the same speaker.}
	\label{fig:sts}
\vspace*{-6mm}
\end{figure}

%% resent research on streaming multi-speaker ASR and remaining gaps
In parallel, researchers also investigated multi-speaker ASR performance under streaming conditions, which is crucial for applications with minimal latency. In \cite{SklyarPiunovaLiu_streaming_rnnt_icassp2021} and \cite{Lu2021}, two conceptually similar streaming multi-speaker ASR systems, \ac{MS-RNN-T} and \ac{SURT} were proposed simultaneously to enable time-synchronous decoding of partially overlapping speech from 2 speakers. Both models are based on \ac{RNN-T} model \cite{graves2012sequence} with implicit speech separation in the encoder and multiple decoding threads (one for each speaker). These approaches were later extended to multi-turn audio processing for any number of speakers in \cite{SklyarPiunova2022multiturn} and \cite{raj2022continuous}, respectively. \ac{SURT} was additionally extended to perform speaker identification in \cite{lu2021streaming} and endpoint detection in \cite{lu2022endpoint}, with limitation to 2-speaker single-turn audio recordings. Recently, an alternative approach to streaming multi-speaker \ac{ASR}, \ac{t-SOT}, was proposed in \cite{KandaTsot}, which unified single-speaker and multi-speaker model architectures by mixing tokens from all speakers in one sequence and sorting them by the order of their appearance in the audio.

%% contribution of this paper
As reported in \cite{SklyarPiunova2022multiturn}, a naïve inclusion of \emph{change-of-turn} (\texttt{<cot>}) segmentation tag into a streaming multi-speaker \ac{ASR} model results in a severe underestimation of the real number of turns in the audio. In this work we attack this issue by explicit turn boundary modeling with \emph{start-of-turn} (\texttt{<sot>}) and \emph{end-of-turn} (\texttt{<eot>}) tokens in a novel separator-transducer-segmenter (STS) model. We further improve recognition and segmentation accuracy of this model through FastEmit \cite{YuChiuFastEmit2021} and masking loss that penalizes leakage of acoustic information into encoder outputs corresponding to non-active speech frames. Finally, we perform segmentation analysis of STS using start-pointing, end-pointing, first subword and last subword emission latency metrics, and apply end-of-turn emission latency penalty to regularize end-pointing emission latency.

\section{Separator-Transducer-Segmenter}
\label{sec:sep-trans-seg}
\subsection{Separator-Transducer}
\label{sec:sep-trans}
\vspace*{-1mm}
STS model inherits its separator and transducer functionalities from a multi-turn RNN-T (MT-RNN-T) \cite{SklyarPiunova2022multiturn}. MT-RNN-T extends the standard \ac{RNN-T} \cite{graves2012sequence} to overlapping speech recognition with multiple output channels \(N\), where $N$ is a maximum number of simultaneously active speakers in the audio. Such design enforces a switch between output channels at speech overlap only and scales to an arbitrary number of speakers. In this work, we only consider cases with $N=2$, and illustrate the corresponding STS model architecture on Fig. \ref{fig:sts}.

The encoder of STS has a modular structure containing a mixture encoder (\(\operatorname{MixEnc}\)), $N$ separation encoders (\(\operatorname{SepEnc}_n\)) for each output channel \(n \in \{1,...,N\} \) and a recognition encoder (\(\operatorname{RecEnc}\)) with shared parameters between output channels. 
The encoder takes acoustic features \(\mathbf{x}\) as input and produces high-level disentangled acoustic representations \( \mathbf{h}_n\) as output:
\vspace*{-2mm}
\begin{align}
\label{eq:speech-encoder}
\mathbf{h}_n = \operatorname{RecEnc}(\operatorname{SepEnc}_n(\operatorname{MixEnc}(\mathbf{x}))).
\end{align}
\vspace*{-4mm}

In order to associate \(\mathbf{h}_n \) with prediction network outputs for each label sequence \(\mathbf{y}_{n}\) we employ \ac{DAT} method \cite{SklyarPiunovaLiu_streaming_rnnt_icassp2021}, which forces the model to learn to associate its output with the speaker order in the audio. In this case, first separation encoder learns to focus on the very first speaker turn, and the second on the follow-up speaker turn if it exists.
As a result, \ac{DAT} computes RNN-T loss only \(N\) times:
\vspace*{-2mm}
\begin{align}
\mathcal{L}_{RNN-T} &= - \sum_{n} \log P(\mathbf{y}_{n} | \mathbf{h}_n) \label{eq:dat-loss}.
\end{align}

\vspace*{-4mm}
\subsection{Segmenter}
\label{sec:segmenter}
\vspace*{-1mm}
On top of the Separator-Transducer model responsible for multi-speaker speech recognition, here we propose a novel Segmenter functionality to perform segmentation of multi-speaker hypotheses into single-speaker hypotheses and estimate turn boundaries for each speaker turn. To achieve this goal, we introduce a new segmentation modelling strategy in Section \ref{sec:seg-modeling}, and explore various regularization methods to enhance its performance in Sections \ref{sec:fast-emit}, \ref{sec:multi-task} and \ref{sec:penalty}.
\vspace*{-2mm}
\subsubsection{Segmentation modeling}
\label{sec:seg-modeling}
\vspace*{-1mm}
In \cite{SklyarPiunova2022multiturn} \emph{change-of-turn} (\texttt{<cot>}) tag was introduced in-between per-turn target transcriptions for each output channel. It was reported that this approach underestimated the number of turns in the audio. In this work, we revisit this design choice and introduce two separate tags for turn segmentation: \emph{start-of-turn} (\texttt{<sot>}) and \emph{end-of-turn} (\texttt{<eot>}). This approach enables joint start-of-turn and end-of-turn detection and allows interpretation of emission timestamps of these tokens as turn boundaries. In the future, these timestamps can be used for other tasks like speaker diarization (as in \cite{XiaLuWangTurnToDiarize}) or endpoint detection (as in \cite{lu2022endpoint}). 

\vspace*{-2mm}
\subsubsection{FastEmit}
\label{sec:fast-emit}
\vspace*{-1mm}
Since in STS model emission timings of \texttt{<sot>} and \texttt{<eot>} tokens act as turn boundaries for potential future application in downstream tasks, it becomes important to regularize their emission latency. To achieve this goal, we use FastEmit \cite{YuChiuFastEmit2021}, a sequence-level emission regularization method. It encourages predicting non-blank tokens and suppresses blank tokens across the entire sequence based on transducer forward-backward probabilities. %FastEmit applies a “higher learning rate”, controlled by a hyperparameter $\lambda_{FastEmit}$, to the prediction of non-blank token when back-propagating into the STS network.

\vspace*{-2mm}
\subsubsection{Multi-task training with masking loss}
\label{sec:multi-task}
\vspace*{-1mm}
To further regularize segmentation capability of the STS model, during training we expose it to the ground-truth segmentation information that is encoded in speech activity labels. Inspired by the work of \cite{Tripathi_2020}, we employ L2 masking loss to penalize recognition encoder outputs $\mathbf{h}_{n}$ in regions with no active speaker. The model is trained in a multi-task fashion by jointly optimizing \ac{RNN-T} and masking losses:

\begin{align}
\label{eq:masking-loss}
\mathcal{L} = \mathcal{L}_{RNN-T} +  \gamma * \sum_{n} L2(\mathbf{h}_{n} \circ \mathbf{m}_{n}),
\end{align}
where $\mathbf{m}_{n}$ is an inverse binary mask of speech activity for output channel $n$ and $\gamma$ is a weight of masking loss. This approach enforces encoder outputs $\mathbf{h}_{n}$ for each output channel $n$ to be close to 0 in frames where there is no active speaker turn assigned to this output channel.

\vspace*{-2mm}
\subsubsection{End-of-turn latency penalty}
\label{sec:penalty}
\vspace*{-1mm}
Besides FastEmit, we also explore another emission regularization method, dynamic latency penalty, applied specifically to the \texttt{<eot>} token. This approach was originally proposed for \emph{end-of-speech} token emission latency regularization in \cite{BoFastAccurate} and investigated for endpoint detection in multi-speaker \ac{ASR} in \cite{lu2022endpoint}. In our case, for each \texttt{<eot>} token in the target transcription, we apply the following dynamic penalty to the probability of \texttt{<eot>} emission in log domain: 
\begin{align}
\label{eq:latency-penalty}
\log P(\langle eot \rangle | \mathbf{x}_t) - = max(0, \alpha(t-\tau - t_{end}))
\end{align}
where  $\alpha$ is a tunable scale of a late \texttt{<eot>} emission penalty, $\tau$ is a \texttt{<eot>} token frame buffer and $t_{end}$ is a ground-truth end-of-turn frame. This penalty increases over time and enforces timely emission of the \texttt{<eot>} token.

\vspace*{-2mm}
\section{Experimental setup}
\label{sec:exp-results}
\vspace*{-1mm}
\subsection{Task description}
\label{sec:task-description}
\vspace*{-1mm}
We perform experiments with \ac{STS} on LibriCSS dataset proposed in \cite{chen2020continuous} for continuous speech separation.
It contains 10 one-hour long audio sessions with LibriSpeech utterances played back in a room to simulate meetings with 8 speakers, and it is divided into 6 partitions.
0S and 0L partitions exclude overlapped speech but contain short (S) and long (L) silence gaps between speaker turns, respectively.
The remaining partitions represent different overlap ratios from 10\% to 40\%: OV10, OV20, OV30 and OV40. We use Session 0 of this dataset as a development set to tune decoding hyper-parameters and select best checkpoints, while the remaining Sessions 1-9 are used to report performance.

Following previous work in \cite{SklyarPiunova2022multiturn}, we adopt an utterance group evaluation protocol (proposed in \cite{kanda2021slt}) for experiments on this dataset. This evaluation protocol enforces segmentation of the original one-hour long audio sessions into utterance group segments using oracle silence boundary information. It ensures the existence of utterance groups containing only one speaker turn (0S, 0L) and utterance groups containing multiple partially overlapping turns. We are aware of the fact that parallel works on streaming multi-speaker ASR \cite{raj2022continuous,KandaTsot} recently adopted an alternative continuous input evaluation protocol from \cite{chen2020continuous}, and we plan to address this discrepancy in the future work.

\begin{table*}[t]
	\caption{WER and turn counting accuracy benchmarking of the STS model variants against the baseline on LibriCSS.}
	\label{tab:libricss-results}
	\vspace*{-3mm}
	\centering
	\begin{tabular}{ l c c  c c c c c c c c}
		\toprule
		\multirow{ 2}{*}{Model} & \multicolumn{2}{c}{Turn counting accuracy [\%]} & \multicolumn{7}{c}{WER [\%]} \\
		& Overall & $>$ 2 turns & 0L & 0S & OV10 & OV20 & OV30 & OV40 & full \\
		\midrule
		MT-RNN-T \cite{SklyarPiunova2022multiturn}  & 85.6  & 28.0 & 14.7  & 14.8  & 20.7  & 25.3  & 33.2  & 36.4 & 25.3 \\
		\midrule
		STS & 89.0 & 43.0 & 14.8  & 14.7 & 18.1  & 24.0  & 32.6  & 38.8 & 25.0  \\
		\hspace{1mm} +FastEmit & 90.1 & 47.5 & 13.0 & 13.5 & 16.0 & 21.3 & 29.3 & 31.3 & 21.7 \\
		\hspace{2mm} +Masking loss & 90.2 & 50.6 & 13.0 & 14.0 & 15.9 & 18.8 & 28.6 & 30.7 & 21.1 \\
		\bottomrule
	\end{tabular}
\vspace*{-4mm}
\end{table*}

\vspace*{-2mm}
\subsection{Training setup}
\label{sec:training-setup}
\vspace*{-1mm}
The model topology of \ac{STS} closely follows the one established in the previous work on \ac{MS-RNN-T} and \ac{MT-RNN-T} \cite{SklyarPiunovaLiu_streaming_rnnt_icassp2021,SklyarPiunova2022multiturn}.
We use 2 LSTM layers in each recurrent module of the architecture (mixture encoder, 2 separation encoders, recognition encoder, prediction network) with 1024 units in each layer. Layer normalization \cite{ba2016layer} is performed after each LSTM layer in the model architecture.
Output layers in the recognition encoder and the prediction network have 640 units. The joint network has a single feed-forward layer with 512 units.
The output softmax layer has a dimensionality of 2503 which corresponds to the blank label, \texttt{<sot>} token, \texttt{<eot>} token and 2500 wordpieces that represent the most likely subword segmentation from
a unigram word piece model \cite{kudo2018sentencepiece}.

Acoustic features are 64-dimensional log-mel filterbanks with a frame shift of 10ms which are stacked and downsampled by a factor of 3.
We use SpecAugment with LibriFullAdapt policy \cite{Park_2020} for feature augmentation. We use the Adam algorithm \cite{kingma2014adam} with the warm-up, hold and decay schedule proposed in \cite{Park_2019} for the optimization of all models. All experiments with enabled FastEmit are done with $\lambda_{FastEmit}=0.005$.

STS model is pre-trained with a single separation encoder on the LibriSpeech dataset. We use on-the-fly data simulation pipeline developed in \cite{SklyarPiunova2022multiturn} for subsequent training on multi-speaker data. For each simulated example, we sample random number of utterances uniformly from the range $\{1, \dots, 5 \}$, scale them to achieve desired energy ratio (sampled from the range between -5 dB and 5 dB) and convolve with an \ac{AIR} before adding to the mixture. Simulated examples longer than 30 seconds are filtered out to avoid out-of-memory errors in the RNN-T loss.

As an additional optimization on top of the segmentation strategy described in Section \ref{sec:seg-modeling}, we remove \texttt{<sot>} from the first turn and \texttt{<eot>} from the last turn in target transcriptions. We motivate this design choice by the absence of leading and trailing silence segments in our experimental setup, which makes modeling of \texttt{<sot>} and \texttt{<eot>} redundant and arguably detrimental in such cases.

\vspace*{-2mm}
\section{Results}
\vspace*{-1mm}
\label{sec:results}
We report speech recognition and turn counting performance of the STS model variants on the LibriCSS dataset in Table \ref{tab:libricss-results}. We measure speech recognition performance in \ac{ORC WER}\cite{SklyarPiunova2022multiturn}, which effectively factors out the reference-hypothesis pairing errors from the actual word recognition errors. We measure turn counting performance in terms of accuracy of the correct number of turn prediction for 2 cases: overall accuracy and accuracy on utterances with $>2$ turns. The latter is of particular interest for us, since cases with 1 or 2 turns are easily tackled by the model with 2 outputs, and do not require explicit segmentation. We select \ac{MT-RNN-T} with the \emph{change-of-turn} token from \cite{SklyarPiunova2022multiturn} as a baseline in this experiment. For experimental models we consider 3 STS variants: vanilla STS with turn boundary modeling through \texttt{<sot>} and \texttt{<eot>} tokens, STS with FastEmit, and STS trained with both FastEmit and masking loss.

As shown in Table \ref{tab:libricss-results}, STS significantly improves turn counting accuracy by 3.4\% abs. (85.6$\rightarrow$ 89.0). This improvement is especially pronounced on utterances with $>2$ turns, where we report 15\% abs. gain in performance (28.0 $\rightarrow$ 43.0). We observe some fluctuations of speech recognition performance among different data partitions, but WER on the full dataset remains on par. This observation clearly shows the benefit of the proposed segmentation modeling strategy for turn counting performance.

STS model with FastEmit further improves turn counting accuracy for utterances with more than 2 turns by 4.5\% abs., and achieves overall relative \ac{WERR} of 13\%. The latter is attributed to halved deletion rate (from 12\% to 6\%), and shifted ratio between insertion and deletion errors (from 0.18 to 0.48). Evidently, it also helps with more reliable turn count estimation, as the model is less prone to delete the whole turn in the worst-case scenario.

Multi-task training with L2-loss brings an additional boost to the turn counting accuracy on utterances with $>2$ turns, which is improved by 3\% abs. Moreover, due to strong regularization effects, multi-task training leads to rel. \ac{WERR} on LibriCSS partitions with high ratio of overlapped speech, i.e. 12\% on OV20, 2\% on OV30 and 2\% on OV40. To better understand the behaviour of the masking loss, we compare per-frame L2 norms of recognition encoder outputs in regions with and without speech activity. We observe that masking loss changes the ratio between the average per-frame norm in ``active`` and ``non-active`` regions from 1.2 to 5.3. This observation shows that leakage of acoustic information into non-active regions can be detrimental to the model performance, but it can be partially mitigated by the masking loss.

%This improvement comes with a cost of WER increase on non-overlapped partitions 0S and 0L. We noticed that multi-task training is specifically helpful to reduce WER in overlapped regions of audio (45.6 → 43.5 \ac{WERR}). We hypothesize that voice activity signals exposed to the model during training indeed preserve acoustic information in the encoder representations for ``active" frames. One outstanding issue observed is the duplication of token sequences in hypotheses for both output channels, which results into increased INS rate. We attribute this to parallel model decoders being unaware of each other and will address the problem in future work.

%Additionally, we performed analysis of recognition encoder outputs for the baseline model and the candidates that were trained in multi-task fashion with masking L2 loss. Using ground truth voice activity labels we calculated mean output norms in both active and non-active regions per each output channel on LibriCSS dev partition. As displayed in Table \ref{tab:encoder-outputs} multi-task training significantly increased the difference in output means between``active" and ``inactive" regions for both outputs. The ratio between active and non-active mean frame norms changed from 1.2 to 4 for Output 0 and from 1.3 to 10 for Output 1. We believe that this enforced behavior contributes to the improved model turn counting accuracy.

\begin{table*}[h!]
	\caption{Segmentation analysis of the STS model variants on LibriCSS. pX is a X-th percentile of emission latency (EL).}
	\label{tab:seg-analysis-results}
	\vspace*{-3mm}
	\centering
	\begin{tabular}{ l c c  c c c c c c c c c c}
		\toprule
		\multirow{ 3}{*}{Model}  & \multicolumn{12}{c}{Emission latency (EL) [ms]} \\
		& \multicolumn{3}{c}{End-pointing (EP)} & \multicolumn{3}{c}{Last subword (LS)} & \multicolumn{3}{c}{Start-pointing (SP)} & \multicolumn{3}{c}{First subword (FS)} \\
		& Mean & p50 & p90 & Mean & p50 & p90 & Mean & p50 & p90 & Mean & p50 & p90 \\
		%\midrule
		%MT-RNN-T \cite{SklyarPiunova2022multiturn}  & 7382 & 6602 &	13430 & 396 & 60 & 270 & 408 & 309 & 550 & 861 & 610 & 932 \\
		\midrule
		STS & 1428 & 1100 & 2611 & 267 & 60 & 230 & 712 & 386 & 666 & 907 & 603 & 794 \\
		\hspace{1mm} +FastEmit & 1509 & 1100 & 2793 & 359 & 10 & 192 & 479 & 211 & 572 & 787 & 555 & 728
 \\
		\hspace{2mm} +Masking loss & 1288 & 980 & 2711 & 74 & -1 & 145 & 332 & 263 & 537 & 570 & 561 & 730 \\
		\bottomrule
	\end{tabular}
\vspace*{-4mm}
\end{table*}

% \begin{table}[]
% 	\caption{Recognition encoder outputs analysis across model outputs and across regions with different speaker activity labels.}
% 	\label{tab:encoder-outputs}
% 	\vspace*{-3mm}
% 	\centering
% 	\begin{tabular}{ l c c c c }
% 		\toprule
% 		\multirow{ 4}{*}{Model} & \multicolumn{2}{c}{Active frames} & \multicolumn{2}{c}{Non-active frames}\\
% 		 & \multicolumn{2}{c}{mean norm} & \multicolumn{2}{c}{mean norm} \\
% 		 & \multicolumn{2}{c}{(per output)} & \multicolumn{2}{c}{(per output)} \\
% 		\midrule
% 		STS (with FastEmit) & 23 & 21 & 18.9 & 16.7 \\
% 		\hspace{1mm} +Masking loss & 17.3 & 14.9 & 4.3 & 1.5 \\
% 		\bottomrule
% 	\end{tabular}
% \vspace*{-4mm}
% \end{table}

\vspace*{-2mm}
\section{Segmentation analysis}
\subsection{Motivation}
\label{sec:seg-analysis-motivation}
\vspace*{-1mm}
Results in Section \ref{sec:results} demonstrate the benefit of the proposed segmentation modeling strategy for the turn counting accuracy. However, in the realistic scenario, we are not only interested in the correct prediction of the number of turns in the audio, but also in the turn boundary estimation, i.e. prediction of start-of-turn and end-of-turn timestamps. End-of-turn timestamps can be used for endpoint detection, i.e. to close a current turn and propagate its transcription to downstream services such as \ac{NLU}. Both start-of-turn and end-of-turn timestamps can assist speaker diarization in assigning a speaker label to each turn. Therefore, in this section, we propose a methodology for performing comprehensive segmentation analysis of the STS model.
\vspace*{-1mm}
\subsection{Methodology}
\label{sec:seg-analysis-methodology}
To better understand token emission behavior of \texttt{<sot>} and \texttt{<eot>} tokens, we extract emission timings for both output channels of the STS model. For each analysis we pick utterances with $>2$ turns as the remaining cases are trivial for a two-output system like ours. Moreover, we only focus on utterances with correctly estimated number of turns. We consider the following emission latency metrics for this analysis:

\noindent
\textbf{End-pointing emission latency (EP EL) --} difference between ground-truth end-of-turn timestamp and emission timing of the \texttt{<eot>} token. Last turn is omitted.

\noindent
\textbf{Last subword emission latency (LS EL) --} difference between ground-truth end-of-turn timestamp and emission timing of the last subword token in this turn. Last turn is omitted.

\noindent
\textbf{Start-pointing emission latency (SP EL) --}  difference between ground truth start-of-turn timestamp and emission timing of the \texttt{<sot>} token. First turn is omitted.

\noindent
\textbf{First subword emission latency (FS EL) --} difference between ground truth start-of-turn timestamp and emission timing of the first subword token in this turn. First turn is omitted.

\begin{figure}[h]
	\includegraphics[width=\linewidth]{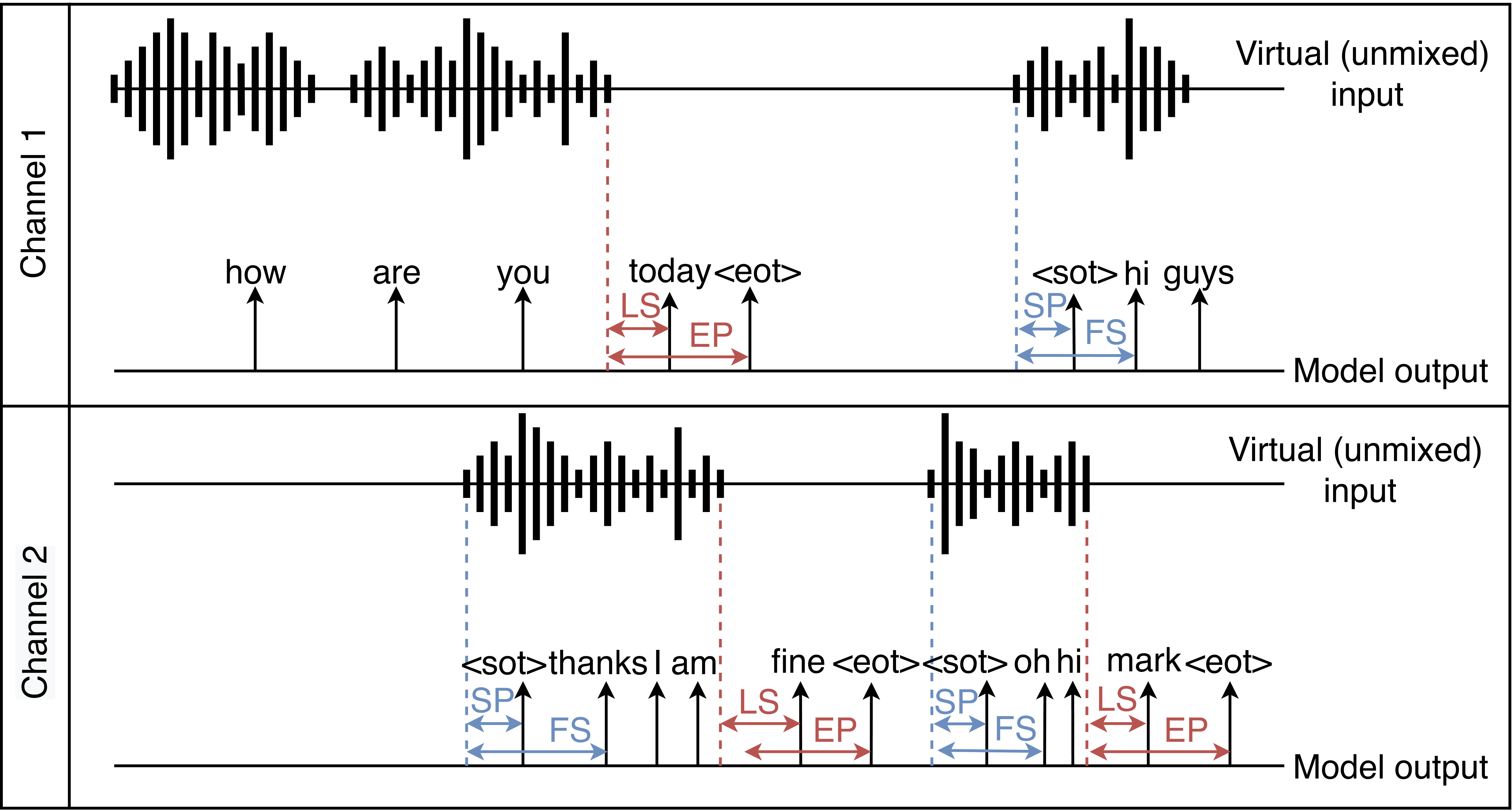}
	\vspace*{0.1mm}
	\vspace*{-6mm}
	\caption{Segmentation analysis example for the proposed STS model. For this analysis, 4 emission latency (EL) metrics are considered: end-pointing (EP), start-pointing (SP), first subword (FS) and last subword (LS).}
	\label{fig:seg-analysis}
\vspace*{-5mm}
\end{figure}

On Fig. \ref{fig:seg-analysis} an example emission latency analysis is depicted. It contains STS model output timestamps for each word and special token as well as ground-truth start-of-turn (blue dashed lines) and end-of-turn (red dashed lines) timestamps for all turns taken into consideration. As seen from this example, LS EL sets a lower bound on EP EL, while FS EL sets an upper bound on SP EL. Difference between SP and FS EL also shows how much audio context STS model needs to open the next turn without predicting the first subword token.

\vspace*{-1mm}
\subsection{Results}
\vspace*{-1mm}
\label{sec:seg-analysis-results}
In Table \ref{tab:seg-analysis-results} we present the results of the segmentation analysis for the STS model variants. Vanilla STS shows the benefit of start-pointing modeling through \texttt{<sot>} token, as SP EL for half of utterances is 35\% smaller than FS EL (386ms vs. 603ms). On a different note, a sizeable gap between LS EL and EP EL reveals a potential caveat of the proposed end-of-turn modeling approach through \texttt{<eot>} token. Average EP EL is around 1.5 sec, which shows that in most cases vanilla STS model significantly delays prediction of \texttt{<eot>}.

FastEmit brings substantial improvements to almost all considered latency metrics. SP EL p50 is almost halved (386ms $\rightarrow$ 211ms), while LS EL p50 is improved by 50ms (60ms $\rightarrow$ 10ms). However, FastEmit does not have an expected emission regularization impact on the \texttt{<eot>} token. This observation is in-line with what was reported in \cite{lu2022endpoint} in the context of \emph{end-of-speech} modeling for multi-speaker ASR, and it motivates us to experiment with a dedicated end-of-turn latency penalty in Section \ref{sec:end-of-turn-latency-penalty-results}. Interestingly, multi-task training with masking loss effectively stabilizes emission latency distribution, which manifests itself in almost 5-fold LS EL reduction (359ms $\rightarrow$ 74ms).
\vspace*{-4mm}
\begin{table}[H]
	\caption{Impact of end-of-turn latency penalty frame buffer $\tau$ on end-pointing emission latency (EL) and WER. pX is a X-th percentile of EL.}
	\label{tab:end-of-turn-latency-penalty-results}
	\centering
	\vspace*{-3mm}
	\setlength{\tabcolsep}{4pt}
	\begin{tabular}{c c c c c c c c}
		\toprule
		\multirow{2}{*}{$\tau$}  & \multicolumn{6}{c}{End-pointing EL [ms]}   & \multirow{2}{*}{\makecell{WER\\ {[\%]}}}\\
	    & Mean & p50 & p60 & p70 & p80 & p90 & \\
		\midrule
		- & 1428 & 1100 & 1440 & 1828 & 2136 & 2611 & 25.0 \\
		\midrule
		3 & 1031 & 14 & 50 & 73 & 134 & 7458 & 27.6  \\
		5 & 1322 & 80 & 100 & 140 & 204 & 7507 & 26.3  \\
	    7 & 1748 & 130 & 158 & 190 & 288 & 9415 & 27.3  \\
		10 & 1058 & 196 & 230 & 260 & 300 & 5388 & 27.7  \\
		\bottomrule
	\end{tabular}
\vspace*{-4mm}
\end{table}

\vspace*{-3mm}
\subsection{Experiments with end-of-turn latency penalty}
\label{sec:end-of-turn-latency-penalty-results}
\vspace*{-1mm}

To specifically improve EP EL, we apply the end-of-turn latency penalty approach described in Section \ref{sec:penalty}. We use vanilla STS as a baseline in this experiment, and apply the end-of-turn latency penalty with a fixed scale $\alpha=1$ and different values of end-of-turn frame buffer $\tau$. As shown in Table \ref{tab:end-of-turn-latency-penalty-results}, it successfully reduces EP EL p50 at least by the order of magnitude for 50-th, 60-th, 70-th and 80-th percentiles. Relaxed end-of-turn frame buffer $\tau$ results in a delayed end-point detection. However, we observe WER and EL EP p90 degradation with all explored $\tau$ values. A more detailed error analysis reveals that they originate from a few end-point detection failures, which lead to the hallucinated hypothesis duplicates from the parallel output channel in affected turns. We tried to address it by combining end-of-turn latency penalty with the best STS model that incorporates both FastEmit and masking loss, with limited success. The major culprit is a numerical instability of the training with both FastEmit and end-of-turn latency penalties, and we plan to address it in the future work.
\vspace*{-2mm}
\section{Conclusion}
\vspace*{-1mm}
In this paper, we proposed Separator-Transducer-Segmenter (STS) model for joint recognition and segmentation of multi-party speech through prediction of turn boundary tokens \texttt{<sot>} and \texttt{<eot>}. It improved turn counting accuracy by 15\% abs. on partially overlapping LibriCSS utterances with $>2$ turns and enabled segmentation analysis based on emission latency of these tokens. On top of STS, three additional modeling changes were explored: an emission regularization method FastEmit, a multi-task training approach with speech activity signal and end-of-turn emission latency penalty. The former two combined additionally improved turn counting accuracy by 7.6\% abs. on utterances with $>2$ turns and overall WER by 16\% rel., while the latter significantly improved end-pointing emission latency for most turns at the cost of slight WER degradation.

\vfill
\pagebreak

\bibliographystyle{IEEEtran}

\bibliography{paper}

\end{document}